\documentclass[prd,preprintnumbers,nofootinbib]{revtex4} 
\usepackage{amsfonts}
\usepackage{amssymb}
\usepackage{amsmath}

\begin{document}
\preprint{YITP-14-77}
\preprint{IPMU14-0310}
\title{A new quasidilaton theory of massive gravity}
\author{Shinji Mukohyama}
\affiliation{Yukawa Institute for Theoretical Physics, Kyoto University,
Kyoto 606-8502, Japan}
\affiliation{Kavli Institute for the Physics and Mathematics of the
Universe (WPI), Todai Institutes for Advanced Study, The University of
Tokyo, 5-1-5 Kashiwanoha, Kashiwa, Chiba 277-8583, Japan}
\date{\today}

\begin{abstract}
 We present a new quasidilaton theory of Poincare invariant massive
 gravity, based on the recently proposed framework of matter coupling
 that makes it possible for the kinetic energy of the quasidilaton
 scalar to couple to both physical and fiducial metrics
 simultaneously. We find a scaling-type exact solution that expresses a
 self-accelerating de Sitter universe, and then analyze linear
 perturbations around it. It is shown that in a range of parameters all
 physical degrees of freedom have non-vanishing quadratic kinetic terms
 and are stable in the subhorizon limit, while the effective Newton's
 constant for the background is kept positive. 
\end{abstract}

\maketitle

\section{Introduction}

It has been a long standing fundamental question in theoretical physics
whether the graviton, a spin-$2$ field that mediates the gravitational
force, can have a finite mass or not. While Fierz and Pauli's pioneering 
work in 1939~\cite{Fierz:1939ix} found a consistent linear theory of
massive gravity, Boulware and Deser in 1972~\cite{Boulware:1973my}
showed that generic nonlinear extensions of the theory exhibit
ghost-type instability, often called Boulware-Deser (BD) ghost. It took
almost 40 years since then until de Rham, Gabadadze and Tolley (dRGT) in
2010~\cite{deRham:2010ik,deRham:2010kj} finally found a nonlinear
completion of the Fierz and Pauli's theory without the BD ghost.

Despite the recent theoretical progress in massive gravity, it is still
fair to say that cosmology in massive gravity has not been established
yet. In this respect, two no-go results are currently known against
simple realization of viable cosmology in massive gravity. The first one 
forbids the flat Friedmann-Lema\^{i}tre-Robertson-Walker (FLRW)
cosmology in the original dRGT theory~\cite{D'Amico:2011jj}. This no-go
can be avoided either by considering open FLRW cosmology in the original
theory~\cite{Gumrukcuoglu:2011ew} or by slightly extending the theory
with a de Sitter or FLRW fiducial
metric (see \cite{Gumrukcuoglu:2011zh} for self-accelerating FLRW
solutions and \cite{Hassan:2011vm,Fasiello:2012rw,Langlois:2012hk} for
non self-accelerating FLRW solutions with a generalized fiducial
metric). However, the second no-go tells that all homogeneous and
isotropic FLRW solutions in the dRGT theory, either 
in its original form or with a more general fiducial metric, are
unstable~\cite{DeFelice:2012mx}. There then seem (at least) three
possible options to go around the second no-go: (i) to relax either
homogeneity~\cite{D'Amico:2011jj} or
isotropy~\cite{Gumrukcuoglu:2012aa,DeFelice:2013awa} of the background
solution; (ii) to extend the theory either by introducing extra
degree(s) of
freedom~\cite{D'Amico:2012zv,Huang:2012pe} or by abandoning the
direct connection with the Fierz and Pauli's
theory~\cite{Comelli:2013txa,Langlois:2014jba,deRham:2014gla}; or (iii)
to change the way how matter fields couple to
gravity~\cite{deRham:2014naa,Gumrukcuoglu:2014xba}.

The quasidilaton theory~\cite{D'Amico:2012zv} introduces an extra scalar
degree of freedom to the dRGT theory and thus falls into the category
(ii). There exists a scaling-type solution that expresses a
self-accelerating de Sitter universe in the flat FLRW chart. However,
the scaling solution in the original theory turned out to be
unstable~\cite{Gumrukcuoglu:2013nza,D'Amico:2013kya}. 
(See \cite{Gabadadze:2014kaa} for another type of self-accelerating
solution in the decoupling limit.) Fortunately, the scaling solution can
be stabilized in a range of parameters by introducing a new coupling
constant corresponding to the amount of disformal transformation to the
fiducial metric~\cite{DeFelice:2013tsa}. For the minimal model of this
type, stability of cosmological evolution in the presence of matter 
fields was recently studied in \cite{Motohashi:2014una}. The theory can
be further generalized as in \cite{DeFelice:2013dua}, allowing for a
larger set of parameters.

In the present paper we shall propose yet another extension of the
quasidilaton theory of massive gravity, motivated by the new matter
coupling~\cite{deRham:2014naa}. The role of the new matter coupling is
to make it possible for the kinetic energy of the quasidilaton scalar to
couple to both physical and fiducial metrics simultaneously.

The rest of the present paper is organized as follows. In
Sec.~\ref{sec:theory} we describe the dRGT theory, the original
quasidilaton theory and the new quasidilaton theory step by step. In
Sec.~\ref{sec:background} we analyze the background equations of motion
with the FLRW ansatz and find an exact scaling-type solution that
expresses a self-accelerating de Sitter universe. This is a continuous
deformation of the same type of solution that was already found in the
original quasidilaton theory. What is interesting is that, unlike the
extension considered in \cite{DeFelice:2013tsa}, properties of the
scaling-type solution depends crucially on a new parameter introduced by 
the extension in the present paper. For example, in the limit of a small
Hubble expansion rate, the effective Newton's constant for the FLRW
background evolution is positive as far as the new parameter is
non-zero, irrespective of other parameters of the theory. In
Sec.~\ref{sec:perturbation} we analyze tensor, vector and scalar
perturbations around the de Sitter solution. Based on the result of the
perturbative analysis, in Sec.~\ref{sec:subhorizon-stability} we study
the stability of subhorizon perturbations around the de Sitter
solution. It is shown that all physical degrees of freedom have finite
quadratic kinetic terms and are stable in a range of parameters while
the effective Newton's constant for the background is positive, even
when the genuine cosmological constant is set to
zero. Sec.~\ref{sec:summary} is devoted to a summary and discussions.

\section{From dRGT to new quasidilaton theory}
\label{sec:theory}

In this section we describe the dRGT theory, the original quasidilaton
theory and the new quasidilaton theory step by step.

\subsection{dRGT}

We begin with describing the dRGT massive gravity
theory~\cite{deRham:2010kj}. In the covariant formulation the theory is
described by a physical metric $g_{\mu\nu}$ and four scalar fields
called St\"uckelberg fields, $\phi^a$ ($a=0,1,2,3$). The theory enjoys
the Poincare symmetry in the St\"uckelberg field space, i.e. the action
is invariant under the following transformation 
\begin{equation}
 \phi^a \to \phi^a + c^a, \quad 
  \phi^a \to \Lambda^a_b\phi^b, 
\end{equation}
where $c^a$ are constants and $\Lambda^a_b$ represents a Lorentz
transformation. Hence the St\"uckelberg fields enter the action only
through the pull-back of the Minkowski metric in the field space to the
spacetime defined as
\begin{equation}
 f_{\mu\nu} = \eta_{ab}
  \partial_{\mu}\phi^a\partial_{\nu}\phi^b,  \quad
  \eta_{ab}={\rm diag}(-1,1,1,1).
\end{equation}
Using the tensor $f_{\mu\nu}$, often called a fiducial metric, it is
convenient to define 
\begin{equation}
{\cal K}^{\mu}_{\nu} = \delta^{\mu}_{\nu}
 - \left(\sqrt{g^{-1}f}\right)^{\mu}_{\ \nu}.
\end{equation} 
The graviton mass terms that describe interactions between the physical
metric and the St\"uckelberg fields are then constructed as
\begin{equation}
 I_{\rm dRGT}[g_{\mu\nu},f_{\mu\nu}] =
  M_{\rm Pl}^2m_g^2\int d^4x\sqrt{-g}\,
  \left[ {\cal L}_2({\cal K})+\alpha_3{\cal L}_3({\cal K})
  +\alpha_4{\cal L}_4({\cal K})\right], 
\end{equation}
where
\begin{eqnarray}
 {\cal L}_2({\cal K}) & = & \frac{1}{2}
  \left(\left[{\cal K}\right]^2-\left[{\cal K}^2\right]\right), \quad
 {\cal L}_3({\cal K}) = \frac{1}{6}
  \left(\left[{\cal K}\right]^3
   -3\left[{\cal K}\right]\left[{\cal K}^2\right]
   +2\left[{\cal K}^3\right]\right), 
  \nonumber\\
 {\cal L}_4({\cal K}) & = & \frac{1}{24}
  \left(\left[{\cal K}\right]^4
   -6\left[{\cal K}\right]^2\left[{\cal K}^2\right]
   +3\left[{\cal K}^2\right]^2
   +8\left[{\cal K}\right]\left[{\cal K}^3\right]
   -6\left[{\cal K}^4\right]\right),
  \label{eqn:def-L234}
\end{eqnarray}
and a square bracket in (\ref{eqn:def-L234}) denotes trace
operation. The theory is free from BD ghost at the fully non-linear
level~\cite{Hassan:2011hr,Hassan:2011ea}. However, it has been a rather
non-trivial task to find stable cosmological solutions.

\subsection{Original quasidilaton}

The quasidilaton theory is an extension of dRGT theory that involves an
extra scalar field, called a quasidilaton. In its covariant formulation
the theory is thus described by the physical metric $g_{\mu\nu}$, the four
St\"uckelberg fields $\phi^a$ and the quasidilaton scalar $\sigma$. In
addition to the Poincare symmetry as described in the previous
subsection, the theory is invariant under the global transformation 
\begin{equation}
 \sigma \to \sigma + \sigma_0, \quad
  \phi^a \to e^{-\sigma_0/M_{\rm Pl}}\phi^a,
  \label{eqn:global-symmetry}
\end{equation}
where $\sigma_0$ is an arbitrary constant. One can construct graviton
mass terms that are invariant under the global transformation by simply 
replacing ${\cal K}^{\mu}_{\nu}$ in the dRGT mass terms with
\begin{equation}
 \bar{\cal K}^{\mu}_{\nu} = \delta^{\mu}_{\nu}
 - e^{\sigma/M_{\rm Pl}}\left(\sqrt{g^{-1}f}\right)^{\mu}_{\ \nu}. 
\end{equation} 
Adding a kinetic term of the quasidilaton scalar, one then obtains
\begin{equation}
 I_{\rm QD}[g_{\mu\nu},f_{\mu\nu},\sigma] =
  M_{\rm Pl}^2m_g^2\int d^4x\sqrt{-g}\,
  \left[ {\cal L}_2(\bar{\cal K})+\alpha_3{\cal L}_3(\bar{\cal K})
  +\alpha_4{\cal L}_4(\bar{\cal K})\right]
  -\frac{\omega}{2}\int d^4x\sqrt{-g}\,
  g^{\mu\nu}\partial_{\mu}\sigma\partial_{\nu}\sigma, 
\end{equation}
where $\omega$ is a dimensionless constant. Thanks to the global
symmetry (\ref{eqn:global-symmetry}), the quasidilaton theory allows for
a scaling-type solution that describes a self-accelerating de Sitter
universe in the flat FLRW chart. While the self-accelerating de Sitter
solution in the original quasidilaton theory is
unstable~\cite{Gumrukcuoglu:2013nza,D'Amico:2013kya}, an extension of
the theory makes the same solution stable in a range of
parameters\cite{DeFelice:2013tsa}.

\subsection{New quasidilaton}
\label{subsec:newquasidilaton}

In the present paper we propose yet another extension of the
quasidilaton theory of Poincare invariant massive gravity. In the
original theory, the kinetic term of the quasidilaton scalar $\sigma$ is
given in terms of the physical metric $g_{\mu\nu}$. In the new theory we
consider the following effective metric to construct the kinetic term of
the quasidilaton scalar. 
\begin{equation}
 g^{\rm eff}_{\mu\nu} = g_{\mu\nu} 
  + 2\beta e^{\sigma/M_{\rm Pl}} g_{\mu\rho}
  \left(\sqrt{g^{-1}f}\right)^{\rho}_{\ \nu}
  + \beta^2 e^{2\sigma/M_{\rm Pl}} f_{\mu\nu}, 
\end{equation} 
where $\beta$ is a dimensionless constant. This is a simple extension of
the effective metric proposed by \cite{deRham:2014naa}. Hereafter, it is
assumed that $\beta$ is non-negative in order to avoid signature change
of the effective metric. It is evident that this effective metric
respects the global quasidilaton symmetry
(\ref{eqn:global-symmetry}). We thus propose the action of the new
quasidilaton theory as 
\begin{equation}
 I_{\rm NQD}[g_{\mu\nu},f_{\mu\nu},\sigma] =
  M_{\rm Pl}^2m_g^2\int d^4x\sqrt{-g}\,
  \left[ {\cal L}_2(\bar{\cal K})+\alpha_3{\cal L}_3(\bar{\cal K})
  +\alpha_4{\cal L}_4(\bar{\cal K})\right]
  -\frac{\omega}{2}\int d^4x\sqrt{-g_{\rm eff}}\,
  g_{\rm eff}^{\mu\nu}\partial_{\mu}\sigma\partial_{\nu}\sigma, 
\end{equation}
where $g_{\rm eff}$ and $g_{\rm eff}^{\mu\nu}$ are the determinant and
the inverse of $g^{\rm eff}_{\mu\nu}$. The new quasidilaton theory is
thus parameterized by ($m_g$, $\alpha_3$, $\alpha_4$, $\omega$,
$\beta$). Adding the Einstein-Hilbert action, the total action is then 
\begin{equation}
 I_{\rm tot} = I_{\rm EH} + I_{\rm NQD}, \quad
 I_{\rm EH} = \frac{M_{\rm Pl}^2}{2}\int d^4x\sqrt{-g}\, (R-2\Lambda). 
 \label{eqn:Itot}
\end{equation}

\section{de Sitter background}
\label{sec:background}

We consider a flat FLRW ansatz 
\begin{equation}
 g_{\mu\nu}dx^{\mu\nu} =
  -N(t)^2dt^2 + a(t)^2\delta_{ij}dx^idx^j, \quad
 \phi^0 = f(t), \quad \phi^i = a_0x^i, \quad 
  \sigma = \bar{\sigma}(t),
\end{equation}
where $a_0$ is a constant. The fiducial metric and the effective metric
are then 
\begin{eqnarray}
 f_{\mu\nu}dx^{\mu}dx^{\nu} & = & 
  -(\dot{f})^2dt^2 + a_0^2\delta_{ij}dx^idx^j, \nonumber\\
 g^{\rm eff}_{\mu\nu}dx^{\mu}dx^{\nu} & = & 
  -(1+\beta rX)^2N^2dt^2 + (1+\beta X)^2a^2\delta_{ij}dx^idx^j, 
  \label{eqn:f&geff-background}
\end{eqnarray}
where an over-dot represents derivative with respect to $t$ and we have
introduced the following quantities 
\begin{equation}
 X = \frac{e^{\bar{\sigma}/M_{\rm Pl}}a_0}{a}, \quad
  r = \frac{\dot{f}a}{Na_0}. 
\end{equation}

The independent background equations of motion are 
\begin{eqnarray}
 0 & = & \frac{\dot{J}}{N} + 4HJ,   \nonumber\\
 3H^2 & = & \Lambda + m_g^2\rho_X
  + \frac{\omega}{2}\frac{(1+\beta X)^3}{(1+\beta rX)^2}
  \left(\frac{\dot{X}}{NX}+H\right)^2, \nonumber\\
  -\frac{2\dot{H}}{N} & = & 
  (1-r)Xm_g^2J_X
  + \frac{\omega}{2}\frac{(1+\beta X)^2[(1+\beta X)+(1+\beta rX)]}
  {(1+\beta rX)^2}  \left(\frac{\dot{X}}{NX}+H\right)^2, 
\label{eqn:backgroundEOM}
\end{eqnarray} 
where $H=\dot{a}/(Na)$ is the Hubble expansion rate and 
\begin{eqnarray}
 J & = & m_g^2 X(1-X)
  \left[3 + 3(1-X)\alpha_3 + (1-X)^2\alpha_4\right]
  + 
  \frac{\omega}{2}\frac{\beta X(1+\beta X)^3}{(1+\beta rX)^2} 
  \left(\frac{\dot{X}}{NX}+H\right)^2, \nonumber\\
 \rho_X & = &
  (X-1)\left[(X-1)(X-4)\alpha_3+(X-1)^2\alpha_4-3(X-2)\right],
  \nonumber\\
 J_X & = & 
  (X-1)(X-3)\alpha_3 + (X-1)^2\alpha_4 + 3-2X, \nonumber\\
\end{eqnarray}

The first equation in (\ref{eqn:backgroundEOM}) implies that $J$ decays
as $\propto 1/a^4$ as the universe expands. We thus have an attractor de
Sitter solution at $J=0$ as
\begin{equation}
 H = m_g h, \quad X = X_0, \quad r=r_0,
\end{equation} 
where $h$, $X_0$ and $r_0$ are constants satisfying 
\begin{eqnarray}
 X_0(1-X_0)\left[3 + 3(1-X_0)\alpha_3 + (1-X_0)^2\alpha_4\right]
  + \frac{\omega}{2}\frac{\beta X_0(1+\beta X_0)^3}
  {(1+\beta r_0X_0)^2}h^2 & = & 0, \nonumber\\
 -3h^2 + \lambda + \rho_{X0}
  + \frac{\omega}{2}\frac{(1+\beta X_0)^3}{(1+\beta rX_0)^2}h^2
  & = & 0, \nonumber\\
 (1-r_0)X_0 J_{X0}
  + \frac{\omega}{2}
  \frac{(1+\beta X_0)^2[(1+\beta X_0)+(1+\beta r_0X_0)]}
  {(1+\beta rX_0)^2}h^2
  & = & 0.
\end{eqnarray} 
Here, $\rho_{X0}$ and $J_{X0}$ are $\rho_X$ and $J_X$, respectively,
evaluated at $X=X_0$ and we have defined $\lambda=\Lambda/m_g^2$.

By using the set of equations, one can express ($\lambda$, $\alpha_3$,
$\alpha_4$) in terms of ($h^2$, $X_0$, $r_0$) as
\begin{eqnarray}
 \lambda & = & 3h^2 + (1-X_0)^2 + 
  \frac{\omega h^2(1+\beta X_0)^2A_{\lambda}}
  {2(1+\beta r_0X_0)^2(r_0-1)X_0^2}, \nonumber\\
 \alpha_3 & = & \frac{2}{X_0-1} + 
  \frac{\omega h^2(1+\beta X_0)^2A_3}
  {2(1+\beta r_0X_0)^2(r_0-1)X_0^2}, \nonumber\\
 \alpha_4 & = & \frac{3}{(X_0-1)^2} + 
  \frac{\omega h^2(1+\beta X_0)^2A_4}
  {2(1+\beta r_0X_0)^2(r_0-1)X_0^2}, \label{eqn:lambda-alpha3-alpha4}
\end{eqnarray} 
where
\begin{eqnarray}
 A_{\lambda} & = & (1-r_0)X_0^2\beta^2 
  + 2[-r_0X_0^2+(1+r_0)X_0-r_0]X_0\beta
  - (1+r_0)X_0^2 + 4X_0 - 2, \nonumber\\
 A_3 & = & 
  \frac{(1-r_0)X_0^2\beta^2 + [(1+r_0)X_0-2r_0]X_0\beta
  +2(X_0-1)}{(X_0-1)^2}, \nonumber\\
 A_4 & = & 
  \frac{(r_0-1)(X_0-3)X_0^2\beta^2
  +2[(2r_0+1)X_0-3r_0]X_0\beta + 6(X_0-1)}{(X_0-1)^3}.
\end{eqnarray} 
One can then calculate partial derivatives of ($\lambda$, $\alpha_3$,
$\alpha_4$) w.r.t. ($h^2$, $X_0$, $r_0$). By inverting the Jacobian
matrix, one obtains 
\begin{equation}
\left(\frac{\partial h^2}{\partial \lambda}\right)_{\alpha_3, \alpha_4}
 = \frac{1}{3}
 \left[
  1+\frac{c_3\omega^2 h^2}{c_1\omega h^2+c_2}
 \right]^{-1}, \label{eqn:dh2dlambda}
\end{equation}
where
\begin{eqnarray}
 c_1 & = & 
  (1+\beta X_0)
  \left\{
  X_0^5(1-r_0)^3\beta^5
  +X_0^4(1-r_0)\left[(r_0X_0^2+2(-r_0^2-r_0+1)X_0+r_0(4r_0-3)\right]\beta^4
  \right.
  \nonumber\\
 & & 
  -X_0^3\left[(2r_0+1)(r_0^2+2r_0-2)X_0^2
  +2(-3r_0^3-6r_0^2+7r_0-1)X_0+(5r_0^3+4r_0^2-9r_0+3)\right]\beta^3
  \nonumber\\
 & & 
  \left.
  +X_0^2(1-X_0)\left[(10r_0^2+5r_0-6)X_0-(7r_0+2)(2r_0-1)\right]\beta^2
  +3X_0(1-X_0)^2(1-4r_0)\beta-3(1-X_0)^2 \right\},
  \nonumber\\
 c_2 & = & 
  2X_0^5(1-r_0)^3(1-X_0)(1+\beta r_0X_0)^2\beta^2, 
  \nonumber\\
 c_3 & = &
  \frac{1}{2}(1-X_0)^2(1+\beta X_0)^6.
\end{eqnarray} 
This quantity must be positive in order for the Hubble expansion rate to
be an increasing function of the energy density coupling to the physical
metric. In other words, the positivity of this quantity is nothing but
the positivity of the effective Newton's constant for the background
FLRW cosmology. For $\beta=0$, the expression (\ref{eqn:dh2dlambda}) 
reduces to the result known in the original quasidilaton as 
\begin{equation}
\left(\frac{\partial h^2}{\partial \lambda}\right)_{\alpha_3, \alpha_4}
 = \frac{2}{6-\omega},\quad \mbox{ for } \beta=0.
\end{equation}
The positivity of this quantity is incompatible with the stability of
the de Sitter attractor solution in the original quasidilaton theory,
i.e. with $\beta=0$. On the other hand, with $\beta> 0$ (see subsection
\ref{subsec:newquasidilaton} for the reason why we do not consider a
negative $\beta$), we shall see that the positivity of 
$(\partial h^2/\partial\lambda)_{\alpha_3,\alpha_4}$ can be compatible
with the stability of the de Sitter attractor solution. For example, if
we take the Minkowski limit ($h\to 0$) while keeping $\beta$ non-zero
then we reach the following universal value, which is positive: 
\begin{equation}
 \left(\frac{\partial h^2}{\partial \lambda}\right)_{\alpha_3, \alpha_4}
  \to \ \frac{1}{3}, \quad (h\to 0 \mbox{ with } \beta \mbox{ kept
  finite and positive}).
\end{equation}

\section{Perturbations}
\label{sec:perturbation}

In this section we analyze tensor, vector and scalar perturbations
around the de Sitter solution that we described in the previous
section. 

\subsection{Tensor perturbations}

For tensor perturbations
\begin{equation}
 \delta g_{ij} = a^2h^{\rm TT}_{ij}
\end{equation}
with $\delta^{ij}h^{\rm TT}_{ij}=0$ and 
$\delta^{ki}\partial_kh^{\rm TT}_{ij}=0$, we expand the total action
(\ref{eqn:Itot}) up to quadratic order in perturbations. After
decomposing the perturbations into Fourier modes, we obtain the
quadratic Lagrangian as
\begin{equation}
 L_T = \frac{M_{\rm Pl}^2}{8}a^3N
  \left[ \frac{|\dot{h}^{\rm TT}_{ij}|^2}{N^2}
   -\left(\frac{k^2}{a^2}+M_{\rm GW}^2\right)|h^{\rm TT}_{ij}|^2
	      \right], \label{eqn:Ltensor}
\end{equation} 
where
\begin{equation}
 M_{\rm GW}^2 = 
  \left[
   \frac{(1+\beta X_0)(\mu_3\beta^3+\mu_2\beta^2+\mu_1\beta+\mu_0)}
   {(X_0-1)^2(r_0-1)(1+\beta r_0X_0)^2}\omega h^2
   + \frac{X_0^3(r_0-1)}{X_0-1}\right]m_g^2.
\end{equation}
and
\begin{eqnarray}
 \mu_3 & = & -X_0^3(1-r_0)^2, \nonumber\\
 \mu_2 & = & 2X_0^2[r_0X_0^2+(r_0^2-2r_0-1)X_0+r_0(3-2r_0)], \nonumber\\
 \mu_1 & = & X_0[(r_0+1)^2X_0^2-8X_0+6-2r_0^2], \nonumber\\
 \mu_0 & = & 2(X_0-1)(X_0r_0+r_0-2).
\end{eqnarray}

\subsection{Vector perturbations}

For vector perturbations
\begin{equation}
 \delta g_{0i} = aN B^{\rm T}_i, \quad
 \delta g_{ij} = \frac{a^2}{2}
 (\partial_iE^{\rm T}_j+\partial_jE^{\rm T}_i),
\end{equation}
with $\delta^{ij}\partial_i B^{\rm T}_{j}=0$ and
$\delta^{ij}\partial_i E^{\rm T}_{j}=0$, we expand the total action 
(\ref{eqn:Itot}) up to quadratic order in perturbations. We find that
the quadratic action does not depend on time derivatives of 
$B^{\rm T}_i$. After decomposing the perturbations into Fourier modes,
one can then eliminate $B^{\rm T}_i$ by solving its equation of motion
as 
\begin{equation}
 B^{\rm T}_i = 
  \frac{c_V^2\frac{k^2}{a^2}}{c_V^2\frac{k^2}{a^2}+M_{GW}^2}
  \frac{a\dot{E}^{\rm T}_i}{2N},
\end{equation}
where
\begin{equation}
 c_V^2 = \frac{(r_0+1)^2(r_0-1)(1+\beta r_0X_0)}
  {2(1+\beta X_0)(1+r_0+2\beta r_0X_0)}
  \frac{M_{\rm GW}^2}{\omega h^2m_g^2}.
\end{equation}
The reduced quadratic Lagrangian is then 
\begin{equation}
 L_V = \frac{M_{\rm Pl}^2}{16}\int d^4x a^3N
  \frac{k^2M_{\rm GW}^2}
  {c_{\rm V}^2\frac{k^2}{a^2}+M_{\rm GW}^2}
  \left[ \frac{|\dot{E}^{\rm T}_i|^2}{N^2}
   -\left(c_V^2\frac{k^2}{a^2}+M_{\rm GW}^2\right)
   |E^{\rm T}_i|^2 \right], \label{eqn:Lvector}
\end{equation}

\subsection{Scalar perturbations}
\label{subsec:scalarperturbations}

For scalar perturbations
\begin{equation}
 \delta g_{00} = -2N^2\Phi, \quad
 \delta g_{0i} = aN \partial_i B, \quad
 \delta g_{ij} = a^2
 \left[2\delta_{ij}\Psi + 
  \left(\partial_i\partial_j
   -\frac{1}{3}\delta_ij\delta^{kl}\partial_k\partial_l\right)E
 \right],
\end{equation}
and
\begin{equation}
 \delta \sigma = M_{\rm Pl}\sigma_1,
\end{equation}
we expand the total action (\ref{eqn:Itot}) up to quadratic order in
perturbations. We find that the quadratic action does not depend on time
derivatives of $\Phi$ and $B$. After decomposing the perturbations into
Fourier modes, one can then eliminate $\Phi$ and $B$ by solving their
equations of motion. We then change the variables from ($\sigma_1$, $E$,
$\Psi$)  to ($\tilde{\sigma}_1$, $E$, $\Psi$) by 
\begin{equation}
 \sigma_1 = \frac{\Psi}{1+\beta r_0X_0} + \tilde{\sigma}_1,
\end{equation}
to find that $\Psi$ is also non-dynamical, i.e. the quadratic Lagrangian
does not contain time derivatives of $\Psi$. One can thus eliminate
$\Psi$ as well by using its equation of motion. Finally, we obtain the 
reduced quadratic Lagrangian for the two dynamical variables 
($\tilde{\sigma}_1$, $E$) of the form 
\begin{equation}
 L_S = \frac{M_{\rm Pl}^2}{2}Na^3
  \left( \frac{1}{N^2}\dot{y}^TK\dot{y} 
   + \frac{2m_g}{N}\dot{y}^TMy
   - m_g^2y^TVy\right),
\end{equation}
where $K=K^T$, $M=-M^T$ and $V=V^T$ are $2\times 2$ matrices and 
\begin{equation}
 y = 
  \left(
   \begin{array}{c}
    \tilde{\sigma}_1 \\
    \frac{k^2E}{6(1+\beta r_0X_0)}
   \end{array}\right).
\end{equation}

Hereafter, we consider subhorizon modes, i.e. modes with $k/a\gg H$. We 
suppose that $H\sim |m_g|$ up to a factor of order unity, meaning that
subhorizon modes satisfy $k/a\gg |m_g|$ as well, and that
$\beta>0$. (See subsection \ref{subsec:newquasidilaton} for the reason
why we do not consider a negative $\beta$.) It is convenient to
introduce 
\begin{equation}
 \kappa \equiv \frac{k}{m_g a}, \quad |\kappa| \gg 1
  \label{eqn:def-kappa}
\end{equation}
as a bookkeeping parameter. With $H \sim |m_g|$ and $\beta >0$, the
matrices $K$, $M$ and $V$ are expanded as 
\begin{eqnarray}
 K  & = & 
 \left(\begin{array}{c c} 
  K_{11} & K_{12} \\ 
       K_{12} & K_{22} \end{array} \right)
 = K^{(0)}
       \left(\begin{array}{c c} 1 & -1 \\ -1 & 1\end{array} \right)
       + \kappa^{-2}
	\left(\begin{array}{c c} 
	 K^{(-2)}_{11} & K^{(-2)}_{12} \\ 
	      K^{(-2)}_{12} & K^{(-2)}_{22} \end{array} \right)
	+ O(\kappa^{-4}), \nonumber\\
 M   & = & 
  \left(\begin{array}{c c} 
   0 & M_{12} \\ 
	-M_{12} & 0 \end{array} \right)
  = M^{(0)}
	\left(\begin{array}{c c} 0 & 1 \\ -1 & 0\end{array} \right)
	+ O(\kappa^{-2}), \nonumber\\
 V  & = & 
 \left(\begin{array}{c c} 
  V_{11} & V_{12} \\ 
       V_{12} & V_{22} \end{array} \right)
 = \kappa^2 V^{(2)}
       \left(\begin{array}{c c} 1 & -1 \\ -1 & 1\end{array} \right)
       + 
	\left(\begin{array}{c c} 
	 V^{(0)}_{11} & V^{(0)}_{12} \\ 
	      V^{(0)}_{12} & V^{(0)}_{22} \end{array} \right)
	+ O(\kappa^{-2}). 
\end{eqnarray}
The leading-order components are 
\begin{eqnarray}
 K^{(0)} & = & \frac{(1+\beta X_0)^3\omega}{1+\beta r_0X_0},
  \nonumber\\
 M^{(0)} & = & -\frac{3}{2}
  \frac{\beta^2X_0^2(1+\beta X_0)^2(r_0-1)^2\omega h}
  {2X_0^2\beta^2+(2+3r_0-r_0^2)X_0\beta+r_0+1}, 
  \nonumber\\
 V^{(2)} & = & 
  \frac{(1+\beta X_0)^3(1+r_0+2\beta r_0X_0)\omega}
  {2X_0^2\beta^2+(2+3r_0-r_0^2)X_0\beta+r_0+1}.
\end{eqnarray}
The following combination of the sub-leading components will also be
needed for the stability analysis in the next section. 
\begin{eqnarray}
 K^{(-2)} & \equiv & 
 K^{(-2)}_{11} + K^{(-2)}_{22} + 2K^{(-2)}_{12} = 
  \frac{9\beta^2X_0^2(1-r_0)(1+\beta X_0)^3\omega h^2}
  {2X_0^2\beta^2+(2+3r_0-r_0^2)X_0\beta+r_0+1}, \nonumber\\
 V^{(0)} & \equiv & 
 V^{(0)}_{11} + V^{(0)}_{22} + 2V^{(0)}_{12} = 0.
\end{eqnarray}

\section{Subhorizon stability}
\label{sec:subhorizon-stability}

In order to avoid instabilities whose time scales are parametrically
shorter than the cosmological time scale $H^{-1}$, we require that modes
with $k/a\gg H$, i.e. subhorizon modes, be stable. Other types of
instabilities, if exist, would be as slow as the standard Jeans
instability and thus could be harmless. Throughout this section we
assume that $H\sim |m_g|$ and that $\beta>0$. (See subsection
\ref{subsec:newquasidilaton} for the reason why we do not consider a
negative $\beta$.)

\subsection{No-ghost condition}

For scalar perturbations, we impose that both of the two eigenvalues of
the matrix $K$ be positive for subhorizon modes. Since
\begin{equation}
 K_{22} = K^{(0)} + O(\kappa^{-2}), \quad
  \frac{\det K}{K_{22}} =  K^{(-2)}\kappa^{-2}
  + O(\kappa^{-4}),
\end{equation}
where $\kappa$ is defined in (\ref{eqn:def-kappa}), the necessary and
sufficient condition for the positivity of the two eigenvalues in the
subhorizon limit is that 
\begin{equation}
 K^{(0)} > 0, \quad K^{(-2)}\kappa^{-2} > 0. 
\end{equation}

For vector perturbations, we shall see in the next subsection that the
absence of gradient instability for subhorizon modes requires that 
$c_V^2>0$. Under this condition, the coefficient of the kinetic term is
positive for subhorizon modes if and only if
\begin{equation}
 M_{\rm GW}^2 > 0.
\end{equation}

For tensor modes, the coefficient of kinetic term is constant and always
positive.

\subsection{Positivity of sound speed squared}

For scalar modes, the kinetic matrix $K$ is diagonalized by the change
of variables from $y$ to $\tilde{y}$ through 
\begin{equation}
 y = \left(\begin{array}{c c}
      1 & 0 \\ -\frac{K_{12}}{K_{22}} & 1 
	   \end{array} \right) 
 \tilde{y}.
\end{equation}
By employing the ansatz~\footnote{This ansatz is appropriate for
$m_g^2>0$. For $m_g^2<0$, one can simply replace $\kappa^2$ and
$\Omega^2$ by $-\kappa^2$ and $-\Omega^2$, respectively, and then all
results below hold. In particular, (\ref{eqn:soundspeedsquared-scalar}), 
(\ref{eqn:dispersion-vector}) and (\ref{eqn:dispersion-tensor}) are
unchanged by this replacement.} 
\begin{equation}
 \tilde{y} \propto 
  \exp\left( i|m_g| \int \Omega Ndt\right), 
\end{equation}
and neglecting the time dependence of $\Omega$, $a$ and $N$ (we are 
interested in modes with $k/a\gg H$), the equations of motion is reduced
to the dispersion relation 
\begin{equation}
 (\det K) \Omega^4
 - [K_{11}V_{22}+K_{22}V_{11}-2K_{12}V_{12}+4(M_{12})^2]
 \Omega^2 + \det V = 0. 
\end{equation}
It is easy to estimate the order of each coefficient as 
\begin{eqnarray}
 & & \det K = \kappa^{-2} K^{(0)}K^{(-2)} + O(\kappa^{-4}) 
  = O(\kappa^{-2}), \nonumber\\
 & &  K_{11}V_{22}+K_{22}V_{11}-2K_{12}V_{12}+4(M_{12})^2
  = \left[ K^{(0)}V^{(0)}+K^{(-2)}V^{(2)}+4(M^{(0)})^2 \right]
  + O(\kappa^{-2})
  = O(\kappa^0),   \nonumber\\
 & &  \det V = \kappa^2 V^{(2)}V^{(0)} + O(\kappa^{0}) = O(\kappa^{0}), 
\end{eqnarray} 
where we have used $V^{(0)}=0$ to show the last equality. Thus there is 
a pair of positive and negative frequency modes with
$\Omega^2=O(\kappa^0)$, corresponding to a vanishing sound speed. The 
other pair of modes corresponds to 
\begin{equation}
 \Omega^2
  = \frac{K^{(0)}V^{(0)}+K^{(-2)}V^{(2)}+4(M^{(0)})^2}
  {K^{(0)}K^{(-2)}}\kappa^2 + O(\kappa^0).
\end{equation}
Hence, we obtain the sound speed squared for this pair as
\begin{equation}
 c_s^2 = \lim_{\kappa\to\infty} \kappa^{-2}\Omega^2
  = \frac{K^{(0)}V^{(0)}+K^{(-2)}V^{(2)}+4(M^{(0)})^2}
  {K^{(0)}K^{(-2)}}
  = \left(\frac{1+\beta r_0X_0}{1+\beta X_0}\right)^2.
  \label{eqn:soundspeedsquared-scalar}
\end{equation}
This is always positive and thus there is no classical instability for
subhorizon modes. This value of $c_s^2$ corresponds to the speed limit
set by the light cone of the background effective metric 
$g^{\rm eff}_{\mu\nu}$ (see (\ref{eqn:f&geff-background})).

For vector perturbations, from the action (\ref{eqn:Lvector}) one can
easily read off the dispersion relation for modes with $k/a\gg H$ as
\begin{equation}
 \Omega^2 = c_V^2 \kappa^2 + O(\kappa^0).
  \label{eqn:dispersion-vector}
\end{equation} 
Hence the absence of classical instability for subhorizon modes requires
that 
\begin{equation}
 c_V^2 > 0.
\end{equation}

As is clear from the quadratic action (\ref{eqn:Ltensor}), the
subhorizon dispersion relation for tensor perturbations is
\begin{equation}
  \Omega^2 = \kappa^2 + O(\kappa^0).
   \label{eqn:dispersion-tensor}
\end{equation}
Thus tensor subhorizon modes are always classically stable.

\subsection{Subhorizon stability and self-acceleration}

In summary, supposing that $H\sim |m_g|$ and that
$\beta>0$, all subhorizon modes are stable if and only if
\begin{equation}
 K^{(0)} > 0, \quad K^{(-2)}\kappa^{-2} > 0, \quad
  c_V^2 > 0, \quad M_{\rm GW}^2 > 0. \label{eqn:UVstabilitry}
\end{equation} 
In addition to these conditions, we require that the effective Newton's
constant for the FLRW background be positive, i.e. 
\begin{equation}
 \left(\frac{\partial h^2}{\partial \lambda}\right)_{\alpha_3, \alpha_4}
  > 0, \label{eqn:Geff>0}
\end{equation}
where the left hand side was calculated in Sec.~\ref{sec:background} and
the result is shown in (\ref{eqn:dh2dlambda}).

Hereafter, we assume that $m_g^2>0$. Among the five conditions shown
in (\ref{eqn:UVstabilitry}) and (\ref{eqn:Geff>0}), the first three can
be restated as
\begin{equation}
 \omega > 0, \quad r_0 > 2+2\sqrt{2}, \quad
  x_- < X_0\beta < x_+,
\end{equation}
where
\begin{equation}
 x_{\pm} =
  \frac{1}{4}\left[r_0^2-3r_0-2\pm (r_0-1)\sqrt{r_0^2-4r_0-4}\right]. 
\end{equation} 
The remaining two conditions are complicated but can be satisfied
simultaneously in a range of parameters. (See explicit self-accelerating
examples below.)

Under the condition $r_0>2+2\sqrt{2}$, it is easy to show that $x_->0$,
meaning that $\beta=0$ is excluded. This is consistent with the result
of \cite{Gumrukcuoglu:2013nza,D'Amico:2013kya}: in the original
quasidilaton theory ($\beta=0$) subhorizon modes always suffer from
ghost instability if the effective Newton's constant for the FLRW
background evolution is positive. On the other hand, if $\beta$ is
non-zero and is between $x_-/X_0$ and $x_+/X_0$ then subhorizon modes
are stable in a range of parameters.

The subhorizon behavior of the new quasidilaton theory considered in
this paper is quite different from that of the original quasidilaton
theory. This is because in the subhorizon limit, various quantities such
as $\det K$ are dominated by terms that are absent for $\beta=0$, where
$\beta$ is the new parameter that measures the strength of the coupling
of the kinetic energy of the quasidilaton scalar to the fiducial
metric. Hence, the $\beta\to 0$ limit and the subhorizon limit do not
commute. In other words, the subhorizon limit of the new quasidilaton
theory with $\beta>0$ is quite different from that of the original
theory. (See subsection \ref{subsec:newquasidilaton} for the reason why
we do not consider a negative $\beta$.)

So far, we kept the cosmological constant $\Lambda$ (or its
dimensionless version $\lambda=\Lambda/m_g^2$) as a
placeholder for ordinary matter in order to calculate the response of the
Hubble expansion rate to the energy density coupling to the physical
metric $g_{\mu\nu}$. On the other hand, since one of the modern
motivations for massive gravity is to explain the origin of the current
acceleration of the universe, it is favorable if the graviton mass term
(as well as the quasidilaton kinetic action) can hold the de Sitter
expansion without the genuine cosmological constant. For this reason we
set $\lambda=0$ from now on.

By setting $\lambda=0$ in (\ref{eqn:lambda-alpha3-alpha4}), one obtains 
\begin{equation}
 \omega = \frac{2(1+\beta r_0X_0)^2(r_0-1)X_0^2}
  {-(1+\beta X_0)^2A_{\lambda}}\left[3+\frac{(1-X_0)^2}{h^2}\right]. 
\end{equation}
We thus consider the subspace of the parameter space defined by this
relation. This subspace is $4$-dimensional and can be spanned by
($\beta$, $h$, $r_0$, $X_0$). In this subspace there are many examples
that satisfy the all five conditions shown in (\ref{eqn:UVstabilitry})
and (\ref{eqn:Geff>0}). For example, 
\begin{equation}
 \beta = 1, \quad
  h = 1, \quad
  r_0 = 5, \quad
  X_0 = 2, \label{eqn:example1}
\end{equation}
and
\begin{equation}
 \beta = \frac{1}{200}, \quad
  h = 1, \quad
  r_0 = 200, \quad
  X_0 = 2, \label{eqn:example2}
\end{equation}
satisfy all five conditions shown in (\ref{eqn:UVstabilitry}) and
(\ref{eqn:Geff>0}). The corresponding parameters in the action are,
respectively, 
\begin{equation}
 \Lambda = 0, \quad
 \beta = 1, \quad
 \omega = \frac{7744}{387}, \quad
 \alpha_3 = \frac{66}{43}, \quad
 \alpha_4 = \frac{165}{43},
\end{equation}
and 
\begin{equation}
 \Lambda = 0, \quad
 \beta = \frac{1}{200}, \quad
 \omega = \frac{1910400000000}{27542016533}, \quad
 \alpha_3 = \frac{5426534}{2699933}, \quad
 \alpha_4 = \frac{24700201}{8099799}.
\end{equation}
All five conditions are satisfied in neighborhoods of these points, at
least.

\section{Summary and discussions}
\label{sec:summary}

We have presented a new quasidilaton theory of Poincare invariant
massive gravity, based on the recently proposed framework of matter
coupling that makes it possible for the kinetic energy of the
quasidilaton scalar to couple both physical and fiducial metrics. We
have found a scaling-type exact solution that expresses a
self-accelerating de Sitter universe, and then analyzed linear
perturbations around it. We have shown that in a range of parameters all
physical degrees of freedom have non-vanishing quadratic kinetic terms
and are stable in the subhorizon limit, while the effective Newton's
constant for the background is kept positive.

The proposal of the present paper relies on a simple extension of the
new matter coupling in massive gravity that was recently introduced in
\cite{deRham:2014naa}. Based on the analysis in the decoupling limit, it
was argued in \cite{deRham:2014naa,deRham:2014fha} that the BD ghost is
absent up to $\Lambda_3=(M_{\rm Pl}m_g^2)^{1/3}$ but it may show up at 
some higher scale. The mass of the BD ghost is expected to be around
$m_{\rm ghost}\sim m_g^3M_{\rm Pl}^2/(\sqrt{\beta}\dot{\chi}\partial_i\chi)$, where
$\chi$ is a canonical scalar field that couples to the effective
metric~\cite{Claudia-Lavinia,deRham:2014naa}. The mass of the BD ghost
is higher for smaller $\beta$. (This is consistent with the fact that
there is no BD ghost up to arbitrarily high scale at classical level if
$\beta=0$.) Simply replacing $\chi$ with the quasidilaton $\sigma$ and
noticing that $\dot{\sigma}\sim M_{\rm Pl}m_g$ on the self-accelerating
background (we still assume that $H\sim m_g$), we obtain 
$m_{\rm ghost}\sim
(\Lambda_3/\sqrt{\beta})\times(\Lambda_3^2/\partial_i\sigma)$. 
This means that for $\partial_i\sigma$ below $\Lambda_3^2$, the lowest
possible mass of the BD ghost would be $\sim \Lambda_3/\sqrt{\beta}$. 
This can be above $\Lambda_3$ if $\beta$ is small enough, and the BD
ghost can be integrated out. (On the other hand, $\dot{\sigma}$ is above
$\Lambda_3^2$ and thus the self-accelerating solution cannot be
described by the standard $\Lambda_3$-decoupling limit.)

For $\beta$ of order unity or higher, it is expected that the BD ghost
reappears in some ways. In the present paper we have explictly shown
that the would-be BD degree of freedom ($\Psi$ in subsection
\ref{subsec:scalarperturbations}) has a vanishing time kinetic term and
thus non-dynamical at the level of the quadratic action 
for any values of $\beta$ and $k/a$. This may be due to high symmetry of
the FLRW background or for other subtle reasons. It is worth while
investigating this issue in more details. For example, as in
\cite{DeFelice:2012mx,Yamashita:2014fga}, one may consider linear
perturbations around a Bianchi-I background with axisymmetry as a
consistent truncation of nonlinear perturbations around the
self-accelerating de Sitter solution in the flat FLRW chart. The sixth
degree of freedom may or may not show up in the linear perturbations
around the Bianchi-I background. If it does then an important question
is how heavy the mass gap is. If and only if it is heavy enough then one
can safely integrate it out. While we admit that this is a rather
important issue, we consider it as outside the scope of the present
paper and leave it for a future work.

It is also worthwhile investigating more general quasidilaton theories
by combining the proposal in the present paper with extra terms
considered in \cite{DeFelice:2013tsa,DeFelice:2013dua}.

\section*{\bf Acknowledgments}

The author thanks Claudia de Rham and Lavinia Heisenberg for useful
comments. He acknowledges the YITP workshop YITP-T-14-04 ``Relativistic
Cosmology'', where he presented some results of the paper. He is
grateful to organizers and participants of the workshop, including
Antonio De Felice, Kazuya Koyama, Misao Sasaki, Takahiro Tanaka, Atsushi
Taruya and Gianmassimo Tasinato for warm hospitality and stimulating
discussions. This work was supported in part by Grant-in-Aid for
Scientific Research 24540256 and WPI Initiative, MEXT, Japan.

\end{document}